# Investigation of structural, electrical and electrochemical properties of $La_{0.6}Sr_{0.4}Fe_{0.8}Mn_{0.2}O_{3-\delta}$ as an intermediate temperature solid oxide fuel cell cathode

F. Yadollahi Farsani*, M. Jafari, E. Shahsavari, H. Salamati

*Department of Physics, Isfahan University of Technology (IUT), Isfahan 84156-83111, Iran.*

*Corresponding author, email: fatemehy91@gmail.com

**Abstract**

$La_{0.6}Sr_{0.4}Fe_{0.8}Mn_{0.2}O_3$ (LSFM) compound is synthesized by Sol-gel method and evaluated as a cathode material for the intermediate temperature solid oxide fuel cell (IT-SOFC). X-ray diffraction (XRD) indicates that the LSFM has a Rhombohedral structure with R-3c space group symmetry. The XRD patterns reveal very small amount of impurity phase in the LSFM and $Y_2O_3$-Stabilized $ZrO_2$ (YSZ) mixture powders sintered at 600, 700 and 800 °C for a week. The maximum electrical conductivity of LSFM is about 35.35 $S.cm^{-1}$ at 783 °C in the air. The oxygen chemical diffusion coefficients, $D_{Chem}$, are increased from $1.39\times10^{-6}$ up to $7.66\times10^{-6}$ $\Omega.cm^2$. Besides, the oxygen surface exchange coefficients, $k_{Chem}$, are obtained to lie between $2.9\times10^{-3}$ and $8.7\times10^{-3}$ $cm.s^{-1}$ in a temperature range of 600-800 °C. The area-specific resistances (ASRs) of the LSFM symmetrical cell are 7.53, 1.53, 1.13, 0.46 and 0.31 $\Omega.cm^2$ at 600, 650, 700, 750 and 800 °C respectively, and related activation energy, $E_a$, is about 1.23 eV.

**Keywords:** Intermediate temperature solid oxide fuel cells; $La_{0.6}Sr_{0.4}Fe_{0.8}Mn_{0.2}O_{3-\delta}$; X-ray diffraction; Electrical conductivity; Area-specific resistance.

# 1 Introduction

Solid oxide fuel cell (SOFC) is an energy conversion device which directly converts chemical energy of fuel to electrical energy [1]. This type of fuel cell has highly attracted researchers due to its high efficiency as a new energy source for future [2,3]. However, their high-temperature operation ranges (800-1000 °C) have led to several issues, such as longtime mechanical and performance degradations and high maintenance and processing cost [4-6]. Hence, many efforts have been devoted to decrease the operating temperature and introduce promising intermediate temperature SOFCs (600-800 °C) or IT-SOFCs [7,8]. This provides several benefits, including the utilization of SOFCs in transportation, computers and other portable devices and space applications [9]. Furthermore, it allows usage of a broader set of material, less degradation of cell, prolonged lifetime and reduced manufacturing costs [6,10].

Since the oxygen reduction reaction (ORR) has high activation energy, the cathode is an important component of the IT-SOFC. The decreasing temperature converts the cathode into the major source of total voltage losses for the whole system [11]. The main contribution of the total voltage loss of an IT-SOFC (especially at low temperatures) is generated by cathode polarization [12]. To improve the output of IT-SOFCs, it is necessary to understand the factors limiting the cathode performance and optimization of cathode materials [13]. Therefore, it is necessary to have cathode materials that are highly catalytic [12]. One of these solutions is to use a material that has both ionic and electronic conductions which is known as mixed ionic and electronic conductors (MIECs) to reduce the cathode polarization [14]. The oxygen chemical diffusion coefficient ($D_{Chem}$) and the oxygen surface exchange coefficient ($k_{Chem}$) values are correlated with more efficient overall ORR [15,16]. As well as a confident cathode material requires a matched thermal expansion coefficient (TEC) compatibility with the electrolyte, at operating temperatures [17].

Perovskite oxides with the general formula of $ABO_{3-\delta}$ are the most widely benefited and studied system in different cathode materials with mixed conducting properties [18]. They can be partially or totally substituted with different elements in A and/or B site to get different ionic and electronic conductivities,

TECs and catalytic activities [17]. At the A site, a combination of alkaline earth metal (Ca, Sr and Ba) and rare earth metal (La) ions have been introduced while one or several mixed-valence transition metal ions (Ni, Mn, Fe, and Co) have been proposed at the B site too [19]. The mixed-valence transition metal ions provide the required catalytic activity for the ORR [17].

Among these, the Strontium-doped Lanthanum with the general formula $La_{1-x}Sr_xMnO_{3-\delta}$ (LSM) has been a very common material and is still being researched [20,21]. Substitution of Ca or Sr into the $LaMnO_3$ increases the electronic conductivity. Moreover, the doped $LaMnO_3$ has a high electro-catalytic activity for ORR and a reasonable TEC match with the most common electrolyte which has chemical stability and mechanical strength, 8 mole % Yttria-Stabilized Zirconia (YSZ) [17,22]. LSM perovskite oxides exhibits a poor ionic conductor at IT-SOFC. On the other hand, $LaCoO_3$-based perovskite oxides are known to be superior in electrocatalytic activity and electrical conductivity, compared with $LaMnO_3$-based perovskite oxides [22]. $La_{1-x}Sr_xCo_yFe_{1-y}O_{3-\delta}$ (LSCF) perovskites based on $LaCoO_3$ have attracted considerable attention as an alternative material for IT-SOFC cathodes [13,23]. Yet, the LSCF cathodes cannot be used in the IT-SOFCs without a problem because of the formation of strontium zirconate ($SrZrO_3$) when it is directly coated on YSZ electrolytes [8]. Likewise, these cathodes have problems associated with the usage of Co, which is responsible for the high TEC values usually found in these materials [21]. On the other hand, the Co is very expensive and is a harmful/carcinogenic element [24]. Given the limitations of these two kinds of materials for IT-SOFCs, people are still seeking a new cathode material.

In the periodic table, Fe is next to Mn and they have an almost close ionic radius ($Fe^{3+}$, 0.645 Å, and $Mn^{3+}$, 0.650 Å). Thus the Fe can be substituted partly or fully by Mn [25,26]. In addition, a research has represented that the amount of undesired reaction with the YSZ for the $La_{0.6}Sr_{0.4}Fe_{0.8}Mn_{0.2}O_3$ perovskites is less than the $La_{0.6}Sr_{0.4}Fe_{0.8}Co_{0.2}O_3$ perovskites [27]. The Fe substitution for Mn improves the $D_{Chem}$ and $k_{Chem}$ [14]. $La_{1-x}Sr_xFe_{1-y}Mn_yO_3$ has also high catalytic property [28]. $D_{Chem}$ and $k_{Chem}$ values for the $La_{0.6}Sr_{0.4}Fe_{0.8}Mn_{0.2}O_3$ ($1.4\times10^{-6}$ $cm^2.s^{-1}$ and $2.5\times10^{-4}$ $cm.s^{-1}$ at 750 °C [28]) are higher than the $La_{0.6}Sr_{0.4}Fe_{0.8}Co_{0.2}O_3$ ($1.86\times10^{-8}$ $cm^2.s^{-1}$ and $5.37\times10^{-6}$ $cm.s^{-1}$ at 800 °C [29]) and the LSM ($10^{-11}$ $cm^2.s^{-1}$ [30] and $9\times10^{-5}$ cm.s$^{-1}$ [31] at 1000 °C). Hence the $La_{0.6}Sr_{0.4}Fe_{0.8}Mn_{0.2}O_3$ could be a suitable cathode material. Many studies have been conducted on electrical and catalytic properties of $La_{0.6}Sr_{0.4}Fe_{0.8}Mn_{0.2}O_3$, and its chemical compatibility with YSZ has been completely evaluated [26-28, 32,33]; nevertheless, to the best of our knowledge, no information has been reported to fully understand the electrochemical properties of $La_{0.6}Sr_{0.4}Fe_{0.8}Mn_{0.2}O_3$ cathode in contact with YSZ electrolyte.

In this study, we synthesized $La_{0.6}Sr_{0.4}Fe_{0.8}Mn_{0.2}O_{3-\delta}$ (LSFM) compound as cathode for IT-SOFC. Investigations were addressed to study the structural, electrical and electrochemical properties of this material. The electrical conductivities of the sample were measured in the presence of air; $D_{Chem}$ and $k_{Chem}$ values were determined at 600, 700 and 800 °C in a sudden exchange of Ar atmosphere to the oxygen. The chemical reactivity was evaluated in contact with YSZ. Symmetric cells were fabricated and a detailed analysis of electrochemical response was studied. The microstructure of the symmetric cells was examined through field emission scanning electron microscopy (FESEM) and the manufactured cells were characterized by electrochemical impedance spectroscopy (EIS) in intermediate temperatures at three different oxygen partial pressures. Besides, the distribution of relaxation time (DRT) of the EIS was investigated to provide a model-independent insight into the cathode processes.

## 2 Experimental

### 2.1 Preparation of LSFM powders

LSFM powders were synthesized by a Sol-gel method. $La(NO_3)_3.6H_2O$ (99%), $Sr(NO_3)_3$ (99%), $Mn(NO_3)_2.4H_2O$ (98.5%), and $Fe(NO_3)_3.9H_2O$ (99%), (all from Merck, Germany) were dissolved into deionized water in a stoichiometric ratio. Ethylene Diamine Tetra Acetic Acid (EDTA, 99.4%, Merck, Germany) and a few drops of Ammonia were dissolved in deionized water. Precursors solution were added to a de-counter on a magnetic stirrer, and finally, Citric Acid (99.5%) which dissolved in deionized water was added. The precursors, EDTA, and citric acid solutions were mixed with a molar ratio of 1.0, 1.0 and 1.5, respectively. The PH value was adjusted to around 7 by adding some Ammonia. The temperature of the resulted solution was increased to 90 °C with a 2 °C.min$^{-1}$ heating rate and was maintained for another 12 hours to obtain the transparent gel. The products temperature was increased up to 350 °C with a 2 °C.min$^{-1}$ heating rate and was held at that temperature for about 48 hours to complete the self-combustion step. Finally, the remained ash was grounded using a mortar, and then calcined for 8 hours at 800 °C in an oven in ambient air with a heating rate of 3 °C.min$^{-1}$ and a subsequent cooling rate of 3 °C.min$^{-1}$ to obtain a black powder.

### 2.2 Symmetrical cell fabrication

The YSZ commercial powders (8 mole % $Y_2O_3$-Stabilized $ZrO_2$, Tosoh, Japan) were used as the electrolyte material. The YSZ powders were pressed into the disk pellets (10 MPa) and sintered at 1380 °C for 30 hours. LSFM powders (1 g) were dispersed in a pre-mixed solution of ethanol (5 g), ethyl cellulose (0.27 g) and olive oil (9.3×10$^{-2}$ g) to form a colloidal suspension using Zirconia balls (diameter ≈ 3 mm) and were ball milled at a rotational speed of 250 rpm for 24 hours to form a homogeneous slurry. The slurry was coated using a simple compressed air painter on both sides of the prepared YSZ pellet. The effective cathode area was about 0.63 cm$^2$. The fabricated symmetric cells (LSFM/YSZ/LSFM) were finally sintered at 800 °C for 8 hours in the air. The silver paste was painted to both sides of the symmetric cell as a current collector.

### 2.3 Structural and morphological characterization

The phase structure of the as-prepared LSFM powders was characterized by X-ray diffraction (Philips XPERT) analysis with Cu-Kα (λ=0.1544 nm) radiation with an angle range of 10º≤2θ≤90º and 0.025º step size at room temperature. The resulted data was analyzed using Rietveld refinement by Fullprof software.

To investigate the chemical reactivity between LSFM cathode material and YSZ electrolyte, these two powders were mixed with 1:1 wt. ratio, and were grounded homogeneously with ethanol in a hand-mortar for 1 hour and dried in air. 3 pellets from the final powders were prepared by using uniaxial pressure (30 MPa) and then fired in an oven in ambient air for a weak at 600, 700 and 800 °C. The fired samples were tested by XRD to identify possible reaction products between the LSFM and YSZ.

Scanning electron microscopy (SEM, Philips X130) was used to study the morphologies and average particle size of the LSFM powder. Along with, the microstructure of the symmetrical was examined by field emission scanning electron microscopy (FESEM). Energy-dispersive spectroscopy (EDS) was further performed to identify the elemental composition of the symmetrical cell.

### 2.4  Electrical and electrochemical performance analysis

For electrical conductivity measurement, the LSFM powders were pressed into pellet (diameter= 0.77 cm and thickness=1mm) using uniaxial pressure (50 MPa). The pellet was sintered at 1200 °C for 7 h in air. The electrical resistance of the pellet was measured using a standard four-probe method. Four silver contact were applied on the pellet. A silver wire was attached to each contact with silver paste. The electrical resistance of the sample was measured in air from room temperature up to 900 °C. A K-type thermocouple was used to measure the temperature. Besides, a direct current magnitude of 1 mA was employed for measuring the electrical resistance. The resistivity ($\rho$) and conductivity ($\sigma$) values were calculated using the relation (1) and (2), respectively.

$$\rho = K \frac{V}{I} t \tag{1}$$

$$\sigma = \frac{1}{R} \tag{2}$$

Where $\rho$ is the resistivity ($\Omega$.cm), K is a correction factor (for our sample = 2.2662), V is the measured voltage (V), I is the applied current (A), t is the sample thickness (cm), $\sigma$ is conductivity and R is resistance [34].

Electrical conductivity relaxation (ECR) method was used to study the oxygen transport properties of LSFM. The oxygen chemical diffusion coefficient, $D_{Chem}$, and the oxygen surface exchange coefficients, $k_{Chem}$, measured by the ECR method. A bar with approximate dimensions of 1.6 mm×3.5 mm×10.1 mm was prepared by uniaxial pressure (1.25 MPa). The bar was sintered at 800 °C for 8 hours. The ECR of the sample was measured at 600, 700, and 800 °C. In the beginning, the area around the sample was filled with pure Ar and the system reached and fixed for a while (one hour) at each desired temperature. After stabilization of temperature, the conductivity of the sample was measured during the change of Ar gas to the flow of 1 l/min of pure oxygen. The resulting conductivity variations were recorded as a function of time.

To determine the electrochemical properties, the LSFM/YSZ/LSFM symmetrical cells were tested in a two-point configuration with Ag wires. The EIS data were acquired using an IVIUMSTAT Potentiostat with the frequency range of $10^{-2}$ Hz up to $10^6$ Hz and wave amplitude of 10 mV in the temperature range of 600-800 °C in the air under open-circuit voltage (OCV) condition. All, EIS measurements were also performed at three different oxygen partial pressures of 1, 0.21, and 0.02 atm from 600 to 800 °C.

## 3 Results and discussion

### 3.1 Powder properties and chemical stability

Fig. 1(a) shows the Rietveld refined pattern of the LSFM cathode powders from the XRD data. The LSFM is a single phase with no diffraction peak of the second phase. Moreover, it has a Rhombohedral perovskite structure with the R-3c space group. The Rietveld refinement calculated lattice parameters are a = 5.506 Å, b = 5.506 Å and c = 13.439 Å. The Chi number value (the factor estimates the matching quality of the experimental data with the calculated data in Fullprof software) was placed around 1.02. It is indicated that the obtained data were successfully refined. These results are well comparable to the other reports for LSFM [27,28,32]. Fig. 1(b) demonstrates the room temperature XRD patterns of the LSFM and YSZ pellets (1:1 wt. ratio) after heat treatment at 600, 700 and 800 °C temperatures for one week. Generally, the chemical compatibility between the cathode and the electrolyte is one of the most important aspects of a SOFC material choice. Hereupon, an unwanted chemical reaction may increase the polarization resistance and deteriorate the SOFC performance seriously [12,26,27]. According to Fig. 1(b), no chemical reaction between the LSFM and YSZ can be detected and all the peaks were indexed to LSFM or YSZ. This means that LSFM cathode has good chemical compatibility with YSZ electrolyte in intermediate temperature ranges. However, forming a secondary phase of $SrZrO_3$ are reported by K. Chen et al. between $La_{0.6}Sr_{0.4}Fe_{0.8}Co_{0.2}O_3$ and YSZ at 800 °C [35]. Along with, Ronghui et al. and Kakinuma et al. have indicated good chemical compatibility between LSFM with $La_{0.9}Sr_{0.1}Ga_{0.8}Mg_{0.2}O_3$ and $(Ba_{0.3}Sr_{0.2}La_{0.5})InO_{2.75}$ electrolyte [32,36].

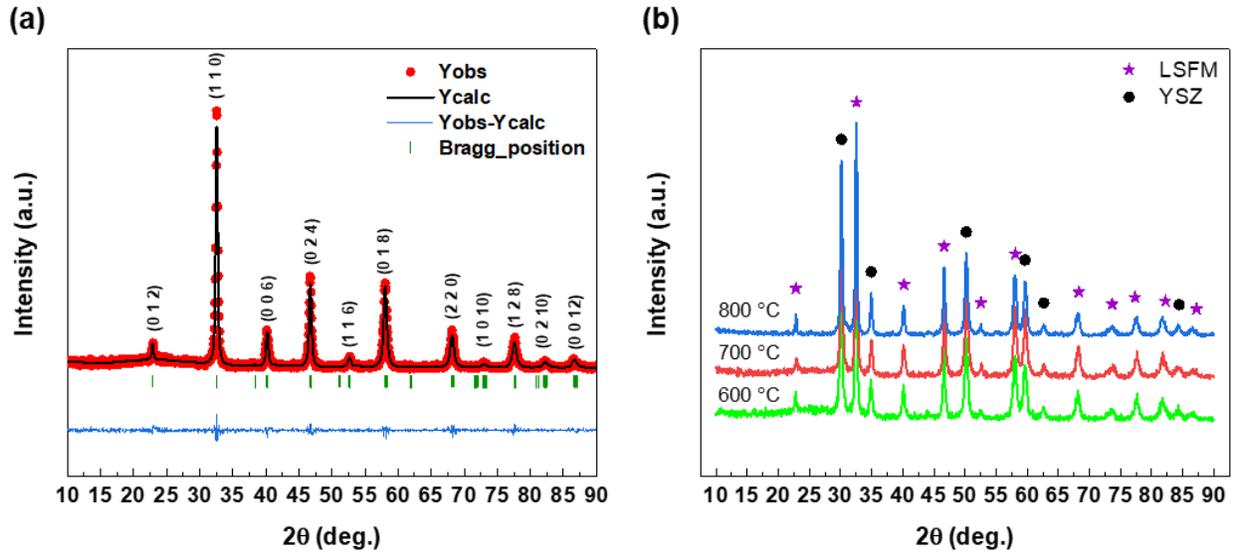

**Fig. 1** – (a) A Rietveld refinement of XRD data for the LSFM powders at room temperature. (b) XRD patterns of the LSFM and YSZ mixture (1:1 wt. ratio), sintered at different temperatures for a week and measured at room temperature.

Fig. 2(a) illustrates the SEM micrograph of the LSFM powder. The powder consists of uniform nano-sized particles and weak agglomerations. Fig. 2(b) displays the estimated particle size distribution chart. The average particle size of the as-prepared LSFM powders was calculated to be about 120±21 nm.

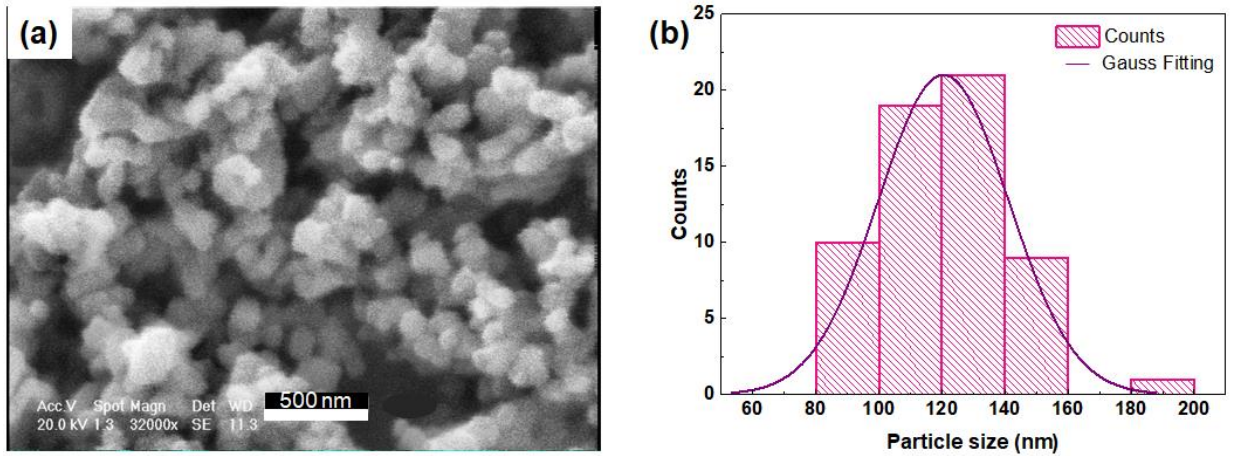

**Fig. 2** – (a) SEM micrograph, and (b) particle size distribution of LSFM powders calcined at 800 °C.

## 3.2 Electrical conductivity and Oxygen transport

Fig. 3(a) demonstrates the temperature dependencies of the electrical conductivity of the LSFM pellet, from room temperature up to 900 °C in ambient air. The electrical conductivity increases with temperature and maximum conductivity was observed at 783 °C. This semi-conductor-like behavior can be attributed to p-type small Polarons hopping mechanism or electronic holes as charge carriers [28]. After 783 °C, the conductivity decreases with increasing temperature (metallic-like behavior). When T ≤ 783 °C, electrons are transmitted via the $Fe^{3+}$-$O^{2-}$-$Fe^{4+}$ and $Mn^{3+}$-$O^{2-}$-$Mn^{4+}$ networks [2]. Though for higher temperatures than 783 °C, the release of the thermally activated lattice oxygen decreases charge carrier concentration which is appeared at high temperatures and oxygen defect is also formed. At high temperatures, oxygen exits from the LSFM and decreases charge carrier (hole) concentration; since $(Fe, Mn)^{4+}$ reduce to $(Fe, Mn)^{3+}$, results in carrier localization decrease in electrical conductivity [2,37-39]. The LSFM had a max conductivity value 35.35 S.cm$^{-1}$, which is higher than the maximum value 16 S.cm$^{-1}$ at 600 °C reported by Ronghui et al. [32]. Also the conductivity at 800 °C (35.24 S.cm$^{-1}$) is lower than that reported by Chung et al. [33]. This observation is probably caused by low density of the pellet. Fig. 3(b) presents the Arrhenius plots of the conductivity versus temperature. The related activation energy ($E_a$) is determined from the slope of the linear part of the Arrhenius plot and according to the relation (3):

$$\sigma = \left(\frac{\sigma_0}{T}\right) \exp\left(\frac{E_a}{k_B T}\right) \tag{3}$$

Where in that σ, $σ_o$, T, and $k_B$ are the lattice conductivity, the pre-exponential constant, absolute temperature and Boltzmann's constant, respectively [17]. The computed $E_a$ for small polarons hopping is about 0.23 eV (30-783 °C) is lower than the reported values for LSFM of 0.25 eV [33].

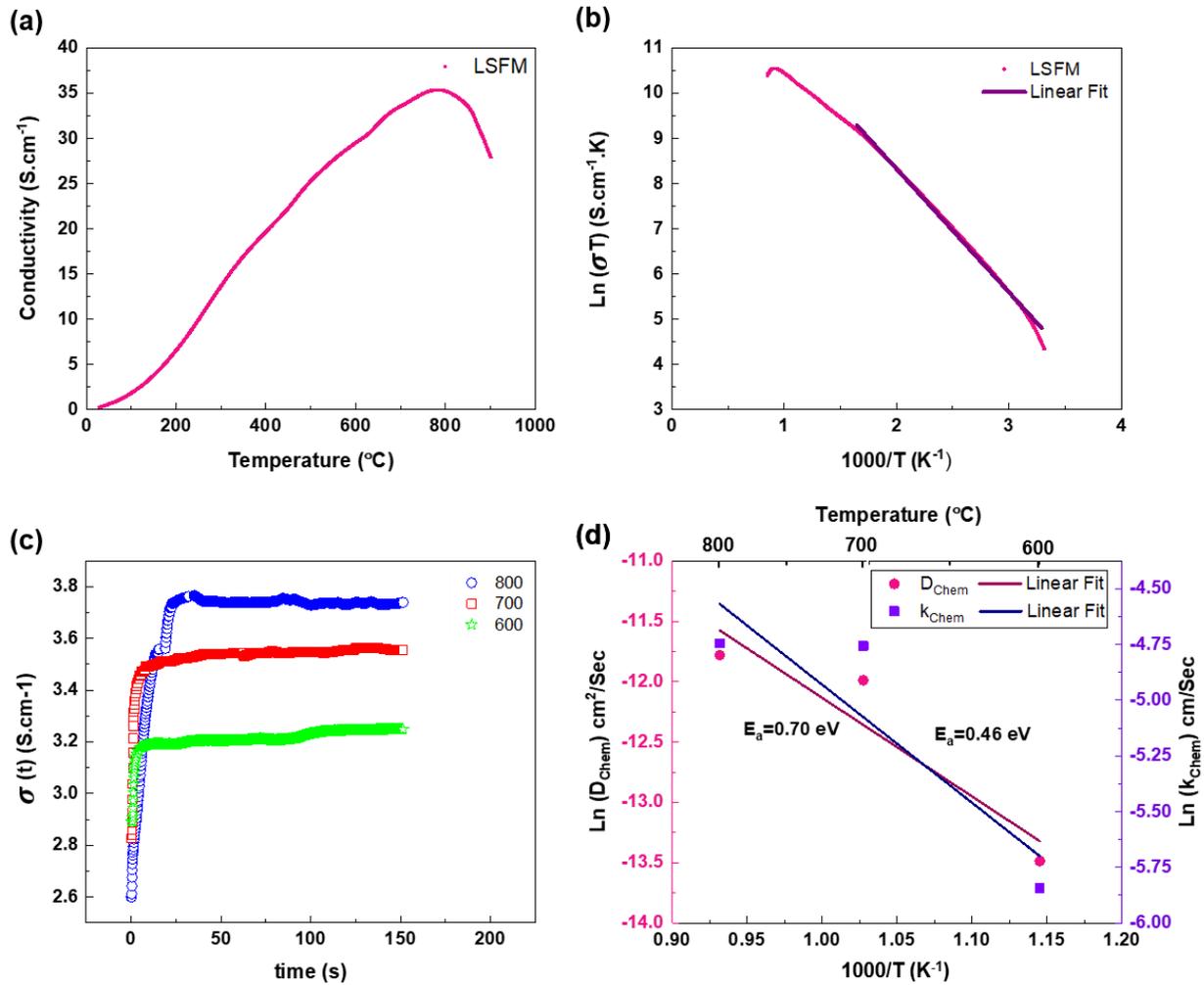

**Fig. 3** – (a) Electrical conductivity of LSFM, (b) related Arrhenius plot of electrical conductivity, (c) electrical conductivity in terms of time during changing atmosphere (from $O_2$ to Ar) at 600, 700 and 800 °C, (d) Arrhenius plot of oxygen diffusion coefficients and oxygen surface exchange coefficients.

Fig. 3(c) shows conductivity variations versus time by the sudden change of the oxygen partial pressures of the sample environment at three different temperatures which have been used for determination of $D_{Chem}$ and $k_{Chem}$. The greater $D_{Chem}$ means lower concentration polarization loss of a cathode. Fig. 3(c) shows the electrical conductivity changes with a sudden change in partial oxygen pressure (The sudden oxygen introduce enhanced the conductivity), and likewise the electrical conductivity increased by increasing temperature. In fact, with the introduction of oxygen into the environment, the conductivity begins to increase and then it's almost stable. The oxygen vacancy concentration increases for temperatures above 300 °C. With the introduction of oxygen, the oxygen is ionized at surface hence oxygen ions occupy oxygen vacancies and become lattice diffused oxygen ions. As a result, the conductivity increases due to an increase in the overlap between O:2p and the Fe/Mn:3d [18,40,41].

The $D_{Chem}$ was specified for three temperatures using a least-squares fit of the experimental data to theoretical equation (4) outlined by Crank et al. The results are presented in Table 1.

$$\frac{\sigma(t)-\sigma(0)}{\sigma(\infty)-\sigma(0)} = 1 - \sum_{n=0}^{\infty} \frac{8}{(2n+1)^2 \pi^2} \exp\left(\frac{-(2n+1)^2 \pi^2 D_{chem} t}{4L^2}\right) \quad (4)$$

Where σ(t) is the conductivity at time t, σ(0) and σ(∞) represent the initial (t=0), and final (t=∞) conductivities, respectively and L is the half of the sample's thickness [18,40].

The $D_{Chem}$ increased with increasing the temperature, and its values are better than other reports. For comparison, Mikkelsen et al. studied the oxygen transport properties of $La_{0.6}Sr_{0.4}Fe_{0.8}Mn_{0.2}O_{3-\delta}$ expressing $D_{Chem}$ values from $1.4 \times 10^{-6}$ to $1.4 \times 10^{-4}$ $cm^2.s^{-1}$ in 750–1000 °C ranges [28]. In addition, Esquirol et al. reported a $D_{Chem}$ value of $4.18 \times 10^{-8}$ $cm^2.s^{-1}$ (T=700 °C) for $La_{0.6}Sr_{0.4}Co_{0.2}Fe_{0.8}O_3/Ce_{0.8}Gd_{0.2}O_2$ composites [42].

| Temperature (°C) | 600 | 700 | 800 |
|---|---|---|---|
| $D_{Chem}$ (cm$^2$.s$^{-1}$) | $1.39 \times 10^{-6}$ | $6.22 \times 10^{-6}$ | $7.66 \times 10^{-6}$ |
| $K_{Chem}$ (cm.s$^{-1}$) | $2.9 \times 10^{-3}$ | $8.6 \times 10^{-3}$ | $8.7 \times 10^{-3}$ |

**Table 1.** Oxygen chemical diffusion coefficients for LSFM at different temperatures.

Fig. 3(d) demonstrates the Arrhenius plots with activation energy ($E_a$) for chemical diffusion and oxygen surface exchange of the LSFM. From the slopes of this plot and equation (5), the $E_a$ for chemical diffusion were calculated about 0.7 eV (68 kJ.mol$^{-1}$). The $E_a$ is less than the values obtained for other cathode reported in the literature. $La_{0.6}Sr_{0.4}Fe_{0.8}Mn_{0.2}O_{3-\delta}$ is reported to have an $E_a$ of 136 kJ.mol$^{-1}$ [28], and for other materials, such as $La_{0.6}Sr_{0.4}Fe_{0.8}Co_{0.2}O_{3-\delta}$ the value is 137.9 kJ.mol$^{-1}$ [18]. As the $E_a$ contains contributions from both migrations of oxygen and oxygen vacancy formation, one possible explanation may be that the low value of $E_a$ for the LSFM reveals the facility of both formations of oxygen vacancy and migration of oxygen in this material [8,28].

$$D = D_0 \exp(\frac{-E_a}{RT}) \tag{5}$$

Where $D_0$ is pre-exponential factor and R is the general constant of gases [18,40].

The conductivity in term of a time constant, τ, can be defined as follows [39,41]:

$$\frac{\sigma(t)-\sigma(0)}{\sigma(\infty)-\sigma(0)} = 1 - \exp(-\frac{t}{\tau}) \tag{6}$$

Where τ is related to the relaxation time. The τ calculated by fitting the curve in Fig. 3(c) in equation (6). Then, the $k_{Chem}$ derived from the following equation:

$$k_{Chem} = \frac{l}{\tau} \tag{7}$$

Where l is the sample's thickness. The $k_{Chem}$ was determined for three temperatures. The results are presented in Table 1. The relation between $k_{Chem}$ and T results the Arrhenius formula:

$$k_{Chem} = k_0 \exp(\frac{-E_a}{RT}) \tag{8}$$

Where $k_0$ is the pre-exponential factor [39,41]. The $E_a$ is obtained about 44.07 kJ.mol$^{-1}$ (0.46 eV). The $k_{Chem}$ value is in good agreement with those reported in the literature for other cathode perovskites. This value is in fact higher than that reported for $La_{0.6}Sr_{0.4}Fe_{0.8}Mn_{0.2}O_{3-\delta}$ (between 2.5×10$^{-4}$ and 1.8×10$^{-2}$ cm.s$^{-1}$ in 750–1000 °C) [28], and $La_{0.6}Sr_{0.4}Fe_{0.8}Co_{0.2}O_{3-\delta}$ (9×10$^{-7}$ cm.s$^{-1}$, at 800°C) [43], (1×10$^{-5}$ cm.s$^{-1}$, at 799 °C) [44]. It is important to note that the oxygen vacancy defects are necessary to induce mixed ionic-electronic conductivity in perovskite materials, favoring desirable ORR for cathode operation in IT-SOFCs [15,16,28,43]. Finally, the above results exhibit that the LSFM cathode has high catalytic activity compared to similar cathodes.

## 3.3  Electrochemical Characterization of Symmetrical Cell

### 3.3.1  EIS results

Fig. 4(a) illustrates the EIS plots of the LSFM/YSZ/LSFM symmetrical cell recorded at 600 °C, 700 °C and 800 °C in the air. The electrochemical performance of each cathode is generally governed by ohmic, activation (charge transfer), and concentration (mass transport) polarizations or drops [22]. The real axis intercept at high frequency refers to the ohmic resistance, $R_1$, mainly contributed by the YSZ solid electrolyte (thickness of the electrolyte), Ag contacts (the contact resistance between the probe and the sample), and wires. The difference between the low and high-frequency intercepts on the real axis represents the polarization resistance, $R_P$ (including activation and concentration polarizations) [2,45]. The $R_P$ and $R_1$ decreased with increasing temperature. There are at least two main overlapped semicircles in the EIS data, which indicate that at least two responses are corresponding to the ORR. The smaller semicircle (high frequency) can be correlated to the charge transfer process at interfaces of current collector/cathode and cathode/electrolyte (activation polarization). While the slightly bigger semicircle (low frequency) is associated with oxygen surface exchange reaction and gas-phase diffusion inside and outside of the cathode through its microstructure (concentration polarization) [2,12,22,45].

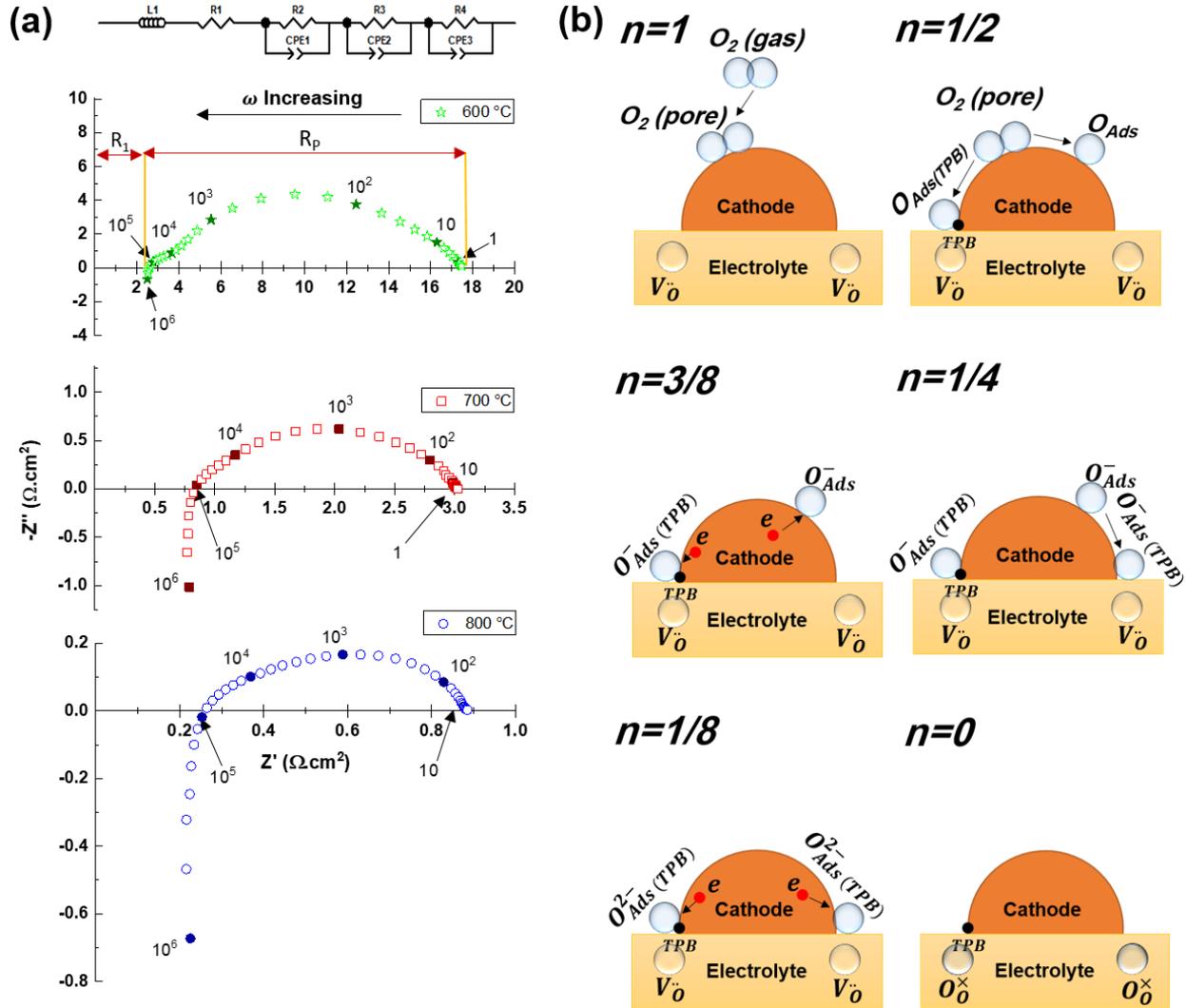

**Fig. 4** – (a) EIS of the LSFM/YSZ/LSFM symmetrical cells at 600, 700, and 800 °C in air and the utilized equivalent circuit for data fitting. (b) Schematic of the ORR reaction mechanism on cathode.

### 3.3.2 EIS results with varying pO$_2$ and DRT results

A distribution of relaxation times (DRT) method was used to study the EIS more precisely and confirming the assumptions [24]. Fig. 5 illustrates the EIS plots for the LSFM symmetrical cell in atmospheres with the different oxygen content of 0.01, 0.21 and 1 atm in the temperature ranges of 600-800 °C. It was found that the polarization resistance decreases as pO$_2$ increases. Information about physio-chemical processes can be obtained by measuring the EIS at different oxygen pressures. This method is a powerful tool to determine the electrode polarization mechanism [24,44].

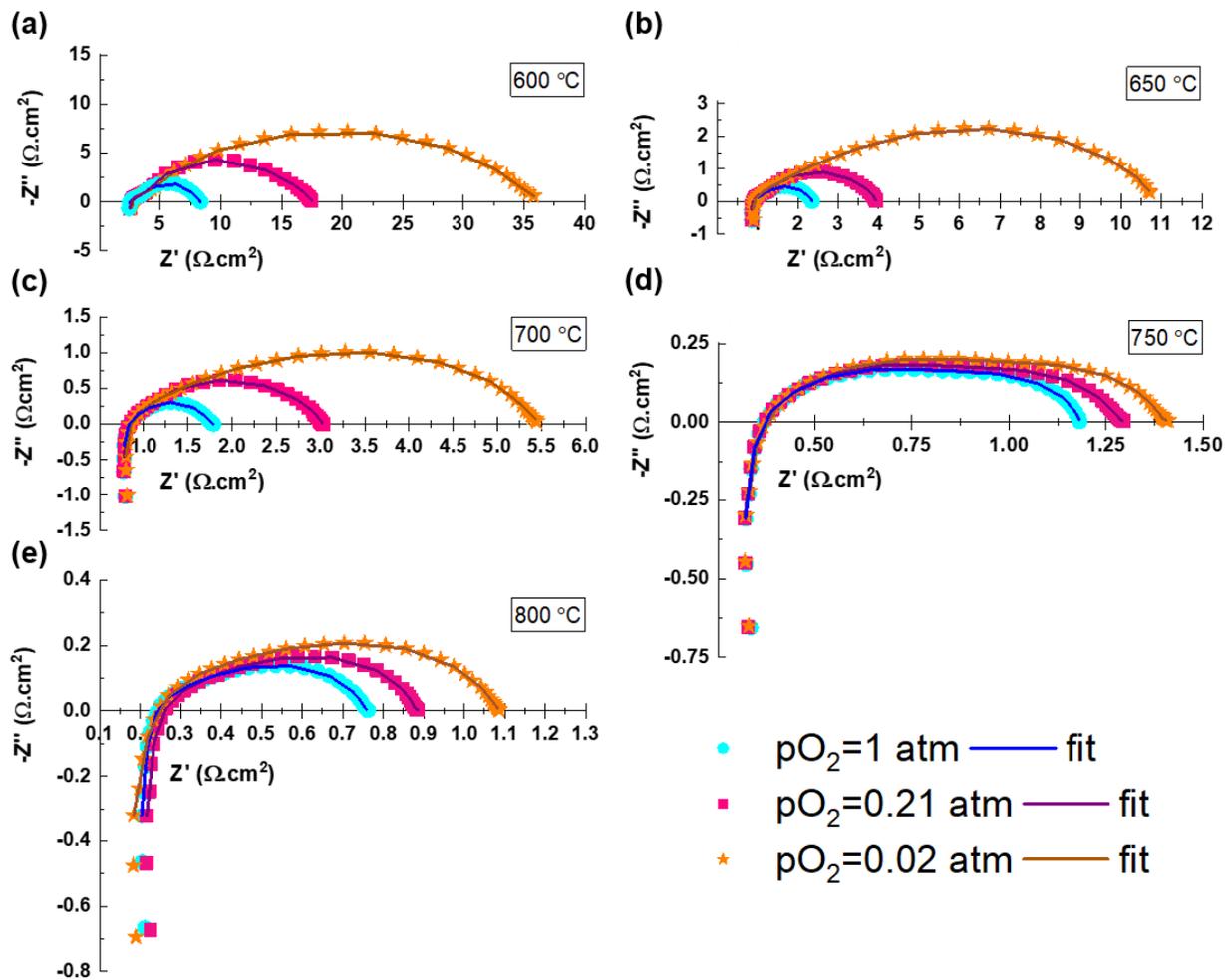

**Fig. 5** – EIS of the symmetrical cell in different oxygen content at (a) 600 °C, (b) 650 °C, (c) 700 °C, (d) 750 °C and (e) 800 °C.

Fig. 6 shows the DRT spectra at 600-800 °C in different partial pressures of $O_2$. Considering acquired EIS data carefully, all data compared of three main peaks including low frequency (LF), intermediate frequency (IF), and high frequency (HF). The EIS spectra can well be fitted by an equivalent circuit ($L_1$-$R_1$-$(R_2CPE_1)$-$(R_3CPE_2)$-$(R_4CPE_3)$) shown in Fig. 4(a). The $L_1$ is the inductance due to the silver current-voltage probes of electrochemical equipment and wires. $R_1$, $R_2$, $R_3$, and $R_4$ are ohmic, HF, IF and LF semicircle resistances, respectively. As well as, $CPE_1$, $CPE_2$, and $CPE_3$ are constant-phase elements of each frequency's semicircle. It should be noted that the other equivalent circuits were also analyzed and appraised, e.g. $L_1$-$R_1$-$(R_2CPE_1)$-$(R_3CPE_2)$-$(GE_1)$ and $L_1$-$R_1$-$(R_2CPE_1)$-$(R_3CPE_2)$-$(Ws_1)$. However, these models caused deterioration of the fits and higher chi-squared values (the chi-squared value for a good fit is <$10^{-5}$). The same equivalent circuit as the one suggested here has been also applied by Zheng et al. [46] for analysis of $La_{0.75}Sr_{0.25}Cr_{0.5}Mn_{0.5}O_3$, and Samat et al. [37] for $La_{0.6}Sr_{0.4}CoO_3$. There is a relatively smaller peak for some spectra at LF, due to their small/negligible contribution to $R_P$, which was removed from the analysis and has not been included in the equivalent circuit modeling. It should be noted that as far as we know, no detailed EIS data analysis has been yet found for the LSFM.

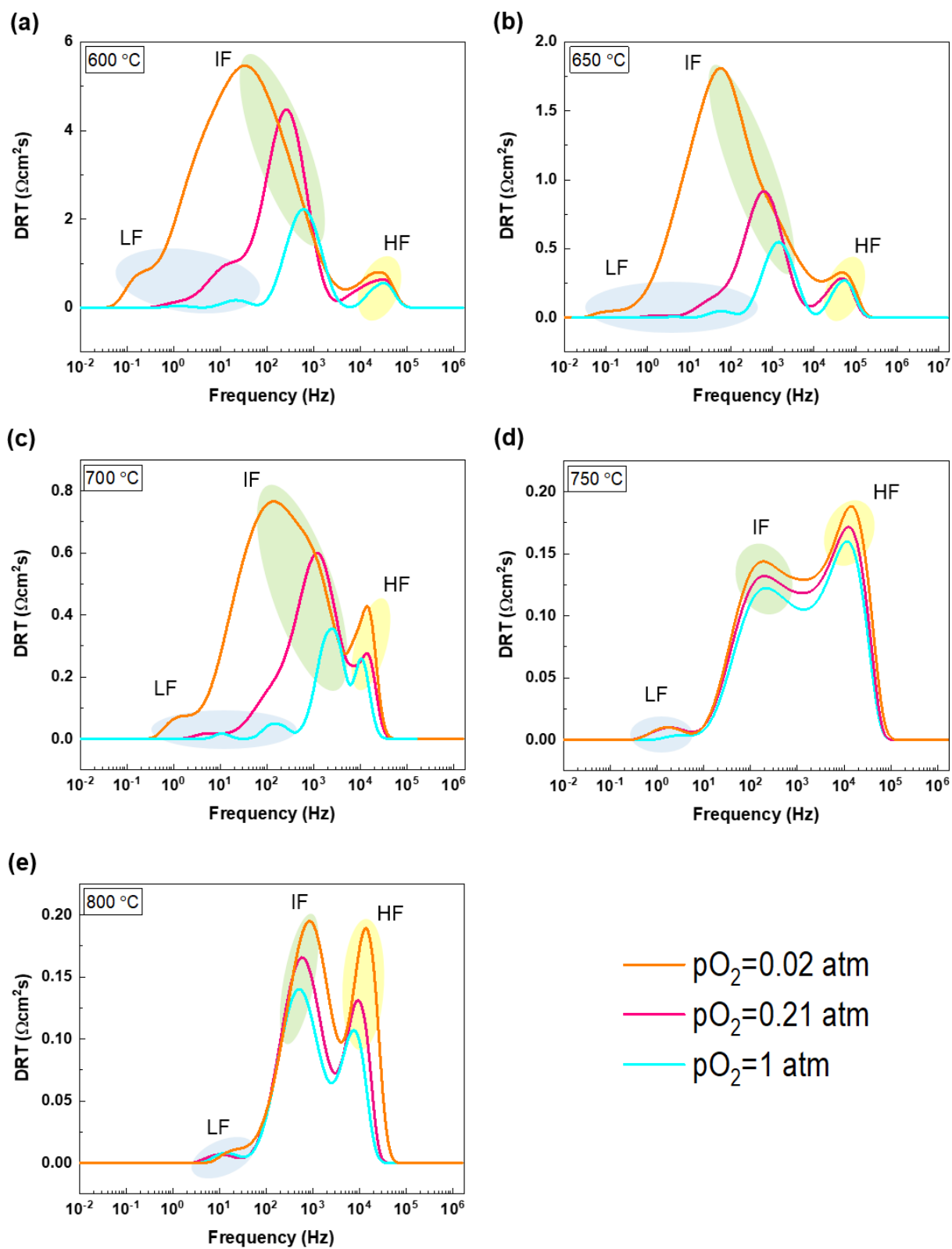

**Fig. 6** – DRT plots of EIS as a function of the oxygen partial pressure at (a) 600 °C, (b) 650 °C, (c) 700 °C, (d) 750 °C and (e) 800 °C.

To get more information for these peaks, the model presented by Heuveln et al. and Bouwmeester et al. were exploited [47]. Based on their dependency on the partial pressures of $O_2$ according to the following equation, the electrochemical processes can be determined [46-48]:

$$R = kP_{O_2}^{-n} \qquad (9)$$

Where k is a constant, R is polarization resistance contributions and the n value can be calculated from the slope of the polarized resistance contributions versus $O_2$ partial pressures (log-log plot). The n values indicate the different possible mechanism [22-24,46-49]. Fig. 4(b) illustrates the meaning of n values. For n=1, oxygen molecule diffusion in the cathode pores is expected,

$$O_2 \,(gas) \rightarrow O_2 \,(pore) \qquad (10)$$

n=0.5 reflects the contribution of atomic oxygen.

$$O_2 + 2(s) \leftrightarrow 2\, O_{Adsorbed}\,(s) \qquad (11)$$

Where (s) is a surface site. n=3/8 is related to electron transfer to the oxygen atom on material surface and/or at the triple-phase boundary (TPB).

$$O_{Adsorbed} + e \rightarrow O^-_{Adsorbed} \qquad (12)$$

n=0.25 assigned to diffusion $O^-_{Adsorbed}$ or $O^{2-}_{Adsorbed}$ along the cathode surface to the TPB region.

$$O^-_{Adsorbed}\,(O^{2-}_{Adsorbed}) \rightarrow O^-_{TPB}\,(O^{2-}_{TPB}) \qquad (13)$$

n=1/8 reflects the reaction of $O^-_{Adsorbed}(O^-_{TPB})$ with the electron to form $O^{2-}_{Adsorbed}\,(O^{2-}_{TPB})$

$$O^-_{Adsorbed}(O^-_{TPB}) + e \rightarrow O^{2-}_{Adsorbed}\,(O^{2-}_{TPB}) \qquad (14)$$

and n=0 is related to the reaction of $O^{2-}$ with oxygen vacancy ($V_{\ddot{O}}$) near the TPB region and diffuses into electrolyte to form lattice oxygen ($O_O^\times$).

$$O^{2-}_{Adsorbed} + V_{\ddot{O}} \rightarrow O_O^\times \qquad (15)$$

Fig. 7(a)-(b)-(c) displays dependencies of resistance on $O_2$ partial pressures in the 600-800 °C ranges. All resistances ($R_2$, $R_3$, and $R_4$) depend on oxygen partial pressure, and the n values for the $pO_2$-dependencies are listed in Table 2. At 600 °C, the n values of $R_2$ and $R_3$ is about 0.17 and 0.28, respectively. These two processes may be associated with adsorption and surface exchange (transport of ionic/electronic defects) [44]. Furthermore, the n value of $R_4$ is about 0.56, which this process was assigned to oxygen dissociative adsorption (involvement of atomic oxygen), dissociation and surface transport [23,50]. According Fig. 6(a), contribution of the IF response is more clearly visible than the other two frequencies response; which is related to the electron transfer process (activation polarizations). For 650 °C, the n value for $R_2$ is 0.20, this result proves that $R_2$ represents the adsorbed oxygen ion diffusion through the cathode surface to the TPB. The $R_3$ and $R_4$ shows n=0.53 and n=0.55 dependency on $pO_2$, which is related to the contribution of atomic oxygen. Fig. 6(b) shows contribution of the IF response is more than LF and HF; which is related to the contribution of atomic oxygen (concentration polarizations). At 700 °C, the related process for each peak was the same as for 600 °C. For 750 °C, from the fitting results, the n value for $R_1$ and $R_2$ is 0.03 and 0.06, which are quite near n = 0. This result shows that $R_1$ and $R_2$ represents the transport process of oxide ions within the bulk cathode and/or from the cathode to the YSZ electrolyte. The calculated n for $R_4$ is 0.12, demonstrating that $R_4$ represents an electron transfer process as shown by eq (14), which corresponds to n = 1/8. The main peak shown in Fig. 6(d) relates to the charge transfer process. At 800 °C, the n values of $R_2$, $R_3$ and $R_4$ is about 0.14, 0.13 and 0.10, respectively; which is associated with the electron transfer (eq (14)). For this temperature also the main peak is relates to the charge transfer process. Therefore it seems that charge transfer process contributes (activation polarizations) the most to the polarization resistance of the LSFM cathode and that is the rate limiting step for ORR. It is also observed that, for all the different oxygen partial pressures, the intensity of the peaks and therefore area of peaks decreases with increasing temperature, this shows both activation polarization resistance and concentration polarization resistance decrease as the temperature increases. Each measured impedance spectrum depend on oxygen transport (both surface processes and diffusion). As mentioned, the concentration polarization contribution

is less than the activation polarization. Because the oxygen transport properties ($D_{Chem}$ and $k_{Chem}$) are an important parameter for the determination of concentration polarization [22].

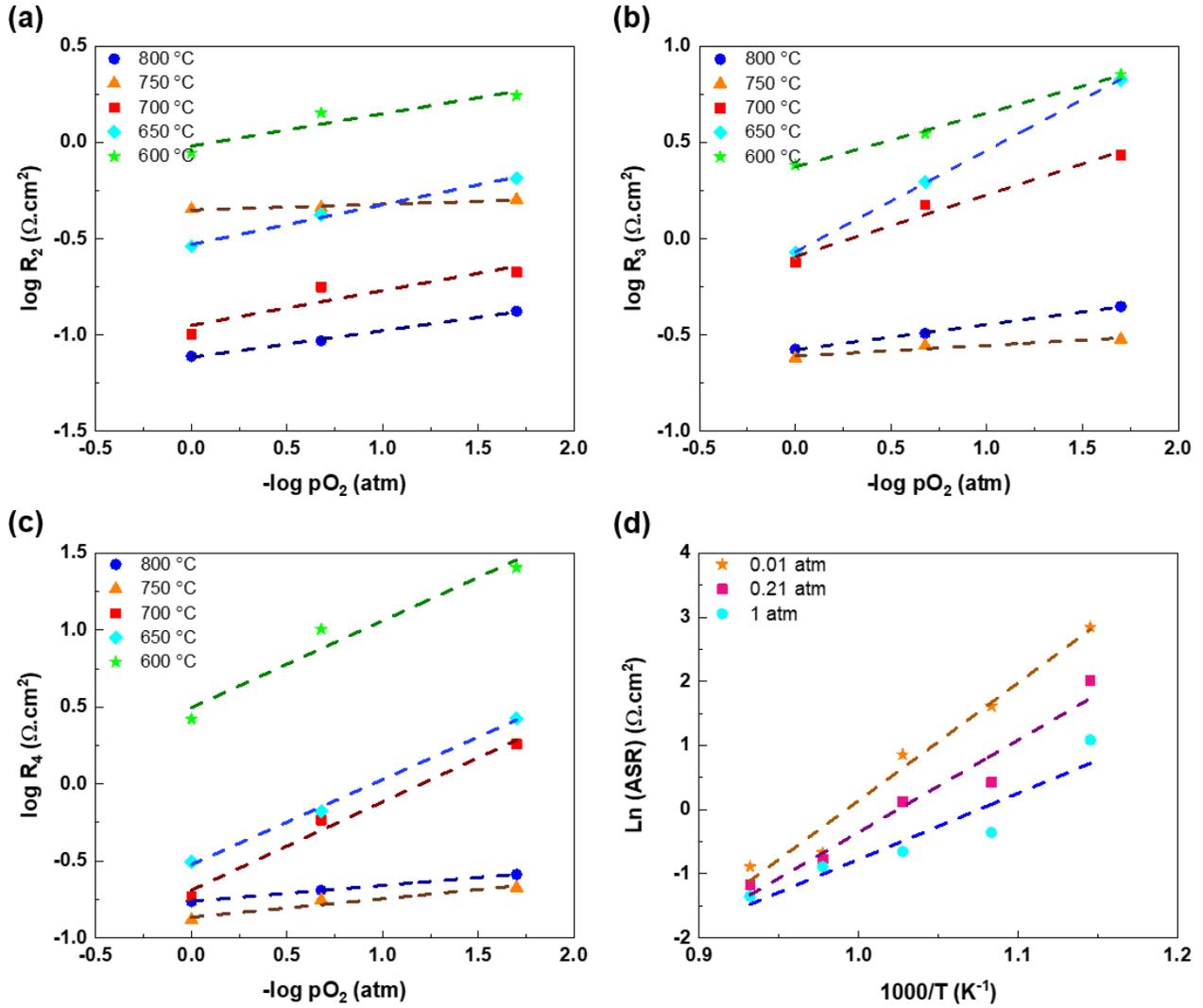

**Fig. 7** – The $pO_2$-dependences of the calculated resistances of (a) $R_2$ (high frequency), (b) $R_3$ (intermediate frequency), and (c) $R_4$ (low frequency) and (d) ASR Arrhenius plots of the LSFM/YSZ/LSFM symmetrical cell in different oxygen partial pressures.

| Temperature (°C) | n value in $(1/R)\alpha P_{O_2}^{-n}$ | | |
|---|---|---|---|
| | $R_2$ | $R_3$ | $R_4$ |
| 600 | 0.17 | 0.28 | 0.56 |
| 650 | 0.21 | 0.53 | 0.55 |
| 700 | 0.18 | 0.32 | 0.57 |
| 750 | 0.03 | 0.06 | 0.12 |
| 800 | 0.14 | 0.13 | 0.10 |

**Table 2.** The n values for $pO_2$–dependences of the EIS fitted Data.

### 3.3.3 ASR results

Using Eq. (16), the area-specific resistance (ASR) was calculated for three different oxygen partial pressures, the results are listed in Table 3.

$$ASR = \frac{R_p \times S}{2} \tag{16}$$

Where $R_p$ is the polarization resistance and S is the cathode surface area.

Fig. 7(d) demonstrates the Arrhenius plots of the ASR of LSFM cathode with different oxygen partial pressures. The $E_a$ of ASRs is calculated from the linear fit slopes using Eq. (17). The $E_a$ of the ASRs is given in Table 3.

$$ASR = A \exp(\frac{E_a}{k_B T}) \tag{17}$$

Where A is constant [12]. The LSFM cathode performance is better than the $La_{0.6}Sr_{0.4}Co_{0.2}Fe_{0.8}O_3$ on YSZ electrolyte with an $R_p$ value of 2.4 $\Omega.cm^2$ in the air at 800 °C when sintered at 1000 °C [45] and better than $LaSrMnO_3$/YSZ composite on YSZ electrolyte with a Rp value of 0.87 $\Omega.cm^2$ in air at 800 °C (sintered at 1180 °C) [51]. Also, the LSFM cathode performance is better than the $(La_{0.8}Sr_{0.2})_{0.95} Fe_{0.2}Mn_{0.8}O_3$ on GDC/YSZ electrolyte with an $R_p$ value of 8.8 Ω in the air at 750 °C [52].

| Oxygen Partial Pressure (atm) | ASR (Ω.cm²) | | | | | ASR-Activation Energy (eV) |
|:---:|:---:|:---:|:---:|:---:|:---:|:---:|
| | 600 °C | 650 °C | 700 °C | 750 °C | 800 °C | |
| 0.01 | 17.14 | 5.02 | 2.36 | 0.51 | 0.41 | 1.58 |
| 0.21 | 7.53 | 1.53 | 1.13 | 0.46 | 0.31 | 1.23 |
| 1 | 2.96 | 0.7 | 0.52 | 0.41 | 0.26 | 0.89 |

**Table 3.** ASR values and activation energies in different oxygen partial pressures.

Fig. 8 shows the cross-sectional FESEM micrographs of the fabricated LSFM/YSZ/LSFM symmetrical cell with two different scales 40 $\mu$m and 10 $\mu$m. The images illustrate that there is a sufficient adherence between the YSZ electrolyte substrate and LSFM cathode film. The LSFM film has a porous microstructure and YSZ electrolyte has a dense structure. The thickness of the cathode and electrolyte layers is around 22.39 and 487.0 μm respectively. In addition, any interlayer indicating possible reactions between two layers (cathode and electrolyte) is not observed by FESEM. Also Fig. 10 indicate the EDX spectra of arbitrarily selected areas on the surface of the symmetrical cell's cross-section and around the interface area. As observed for the position of spot 1, the EDX pattern confirms the presence of La, Sr, Fe, Mn, and O elements at cathode film. Besides, the constituent elements in the position of spot 2, i.e. Zr, Y, and O can be observed from EDX spectra.

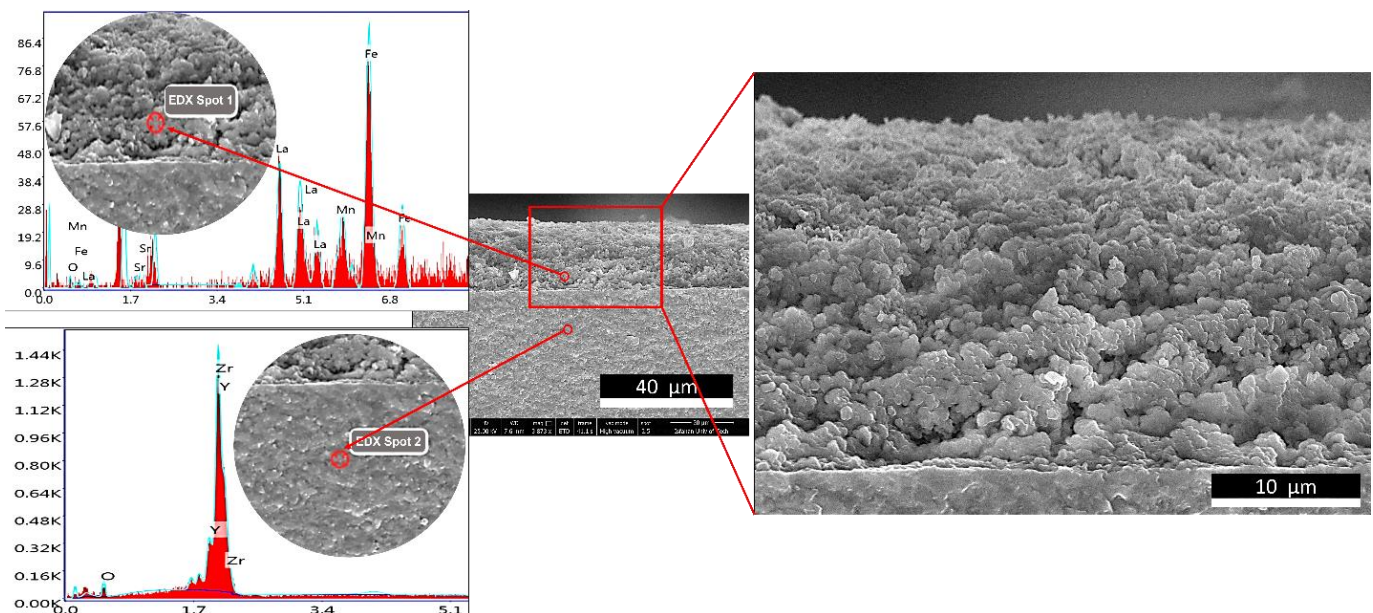

**Fig. 8** – Cross-sectional FESEM of a symmetrical cell with two different magnifications 40 $\mu$m and 10 $\mu$m (The amplification of cathode layer), and EDS patterns at two different spots on the cell's cross-section.

The electrochemical performance of the cathode is dependent on factors such as chemisorption dissociation and reduction reaction of $O_2$ or surface reaction rate ($k_{Chem}$), rate of $O_2$ bulk diffusion ($D_{Chem}$), the electrical conductivity and chemical reactivity at electrode/electrolyte interface. At IT, due to high values of $D_{Chem}$ and $k_{Chem}$ LSFM displayed a better electrocatalytic activity compared to $La_{0.6}Sr_{0.4}Co_{0.8}Fe_{0.2}O_{3-\delta}$ cathode. Also, good values of $D_{Chem}$ and $k_{Chem}$ are evidences of good values of cathode polarization at different temperatures. In view of these properties, it suggested that LSFM is a promising cathode material for IT-SOFC.

## 4 Conclusions

Briefly, $La_{0.6}Sr_{0.4}Fe_{0.8}Mn_{0.2}O_3$ have been synthesized by Sol-gel method, and its structural, electrical and electrochemical properties were investigated as a cathode material for IT-SOFCs. According to the XRD analysis, the LSFM was identified by a Rhombohedral structure with R-3c space group symmetry and no secondary phases were observed. The LSFM cathode has a good chemical compatibility with YSZ electrolyte in operating temperatures. The $D_{Chem}$ for different temperature was in of the order of $10^{-6}$ cm$^{-2}$s$^{-}$ [1] and the activation energy was about 0.7 eV. The k$_{Chem}$ were also measured as 2.9×10$^{-3}$ cm$^2$.s$^{-1}$, 8.6×10$^{-3}$ cm$^2$.s$^{-1}$ and 8.7×10$^{-3}$ cm$^2$.s$^{-1}$ at 600, 700 and 800 °C, respectively. The related activation energy was calculated to be around 44.07 eV. These values indicated that the LSFM has enough electrocatalytic activity towards ORR and also D$_{Chem}$ and k$_{Chem}$ high values are the key factors for improved/acceptable performance of cathode for IT-SOFC. According to the EIS results, the ASR values at 600, 650, 700, 750 and 800 °C were 7.53, 1.53, 1.13, 0.46 and 0.31 Ω.cm$^2$, respectively with an activation energy of about 1.23 eV. Although LSFM presents a slightly high electrical resistance, its great chemical stability and good values of D$_{Chem}$, k$_{Chem}$, and ASRs makes it a suitable choice for being candidate as a cathode material in IT-SOFCs.

**Acknowledgments**

The assistance of S. Hosseini and K. Rahmani is acknowledged.

**References**

[1] L. S. Mahmud, A. Muchtar, M. R. Somalu. "Challenges in fabricating planar solid oxide fuel cells: a review." *Renewable and Sustainable Energy Reviews* 72 (2017) 105-116.

https://doi.org/10.1016/j.rser.2017.01.019

[2] K. P. Padmasree, Ke-Yu Lai, A. F. Fuentes, A. Manthiram. "Electrochemical properties of Sr$_{2.7-x}$Ca$_x$Ln$_{0.3}$Fe$_{2-y}$Co$_y$O$_{7-δ}$ cathode for intermediate-temperature solid oxide fuel cells." *International Journal of Hydrogen Energy* 44, no. 3 (2019) 1896-1904.

https://doi.org/10.1016/j.ijhydene.2018.11.129

[3] T. Suzuki, M. Awano, P. Jasinski, V. Petrovsky, H. U. Anderson. "Composite (La, Sr)MnO$_3$–YSZ cathode for SOFC." *Solid State Ionics* 177, no. 19-25 (2006) 2071-2074.

https://doi.org/10.1016/j.ssi.2005.12.016

[4]    M. Rafique, H. Nawaz, M. S. Rafique, M. B. Tahir, G. Nabi, N. R. Khalid. "Material and method selection for efficient solid oxide fuel cell anode: Recent advancements and reviews." *International Journal of Energy Research* 43, no. 7 (2019) 2423-2446. https://doi.org/10.1002/er.4210

[5]    I. Jang, S. Kim, C. Kim, H. Yoon, T. Song. "Enhancement of oxygen reduction reaction through coating a nano-web-structured $La_{0.6}Sr_{0.4}Co_{0.2}Fe_{0.8}O_{3-\delta}$ thin-film as a cathode/electrolyte interfacial layer for lowering the operating temperature of solid oxide fuel cells." *Journal of Power Sources* 392 (2018) 123-128. https://doi.org/10.1016/j.jpowsour.2018.04.106

[6]    S. C. Singhal. "Solid oxide fuel cells: status, challenges and opportunities." In *Advances in Science and Technology*, Trans Tech Publications, 45.(2006) 1837-1846.

https://doi.org/10.4028/www.scientific.net/AST.45.1837

[7]    O. Yamamoto. "Solid oxide fuel cells: fundamental aspects and prospects." *Electrochimica Acta* 45, no. 15-16 (2000) 2423-2435.

https://doi.org/10.1016/S0013-4686(00)00330-3

[8]    N. Ortiz-Vitoriano, I. R. de Larramendi, S. N. Cook, M. Burriel, A. Aguadero, J. A. Kilner, T. Rojo. "The Formation of Performance Enhancing Pseudo-Composites in the Highly Active $La_{1-x}Ca_xFe_{0.8}Ni_{0.2}O_3$ System for IT-SOFC Application." *Advanced Functional Materials* 23, no. 41 (2013) 5131-5139.

https://doi.org/10.1002/adfm.201300481

[9]    S. A. M. Ali, M. Anwar, N. A. Baharuddin, M. R. Somalu, A. Muchtar. "Enhanced electrochemical performance of LSCF cathode through selection of optimum fabrication parameters." *Journal of Solid State Electrochemistry* 22, no. 1 (2018) 263-273. https://doi.org/10.1007/s10008-017-3754-5

[10] J. H. Zhu, H. Ghezel-Ayagh. "Cathode-side electrical contact and contact materials for solid oxide fuel cell stacking: a review." *International Journal of Hydrogen Energy* 42, no. 38 (2017) 24278-24300. https://doi.org/10.1016/j.ijhydene.2017.08.005

[11] A. Tarancón, M. Burriel, J. Santiso, S. J. Skinner, J. A. Kilner. "Advances in layered oxide cathodes for intermediate temperature solid oxide fuel cells." *Journal of Materials Chemistry* 20, no. 19 (2010) 3799-3813. http://dx.doi.org/10.1039/B922430K

[12] M. Jafari, H. Salamati, M. Zhiani, E. Shahsavari. "Enhancement of an IT-SOFC cathode by introducing YSZ: Electrical and electrochemical properties of $La_{0.6}Ca_{0.4}Fe_{0.8}Ni_{0.2}O_{3-\delta}$-YSZ composites." *International Journal of Hydrogen Energy* 44, no. 3 (2019) 1953-1966.

https://doi.org/10.1016/j.ijhydene.2018.10.151

[13] J. D. Castro-Robles, N. Soltani, J. Á. Chávez-Carvayar. "Structural, morphological and transport properties of nanostructured $La_{1-x}Sr_xCo_{0.2}Fe_{0.8}O_{3-\delta}$ thin films, deposited by ultrasonic spray pyrolysis." *Materials Chemistry and Physics* 225 (2019) 50-54.

https://doi.org/10.1016/j.matchemphys.2018.12.053

[14] M. Petitjean, G. Caboche, E. Siebert, L. Dessemond, L-C. Dufour. "$(La_{0.8}Sr_{0.2})(Mn_{1-y}Fe_y)O_{3\pm\delta}$ oxides for ITSOFC cathode materials?: Electrical and ionic transport properties." *Journal of the European Ceramic Society* 25, no. 12 (2005) 2651-2654. https://doi.org/10.1016/j.jeurceramsoc.2005.03.117

[15] A. Chroneos, B. Yildiz, A. Tarancón, D. Parfitt, J. A. Kilner. "Oxygen diffusion in solid oxide fuel cell cathode and electrolyte materials: mechanistic insights from atomistic simulations." *Energy & Environmental Science* 4, no. 8 (2011) 2774-2789. http://dx.doi.org/10.1039/C0EE00717J

[16] R. Jacobs, T. Mayeshiba, J. Booske, D. Morgan. "Material discovery and design principles for stable, high activity perovskite cathodes for solid oxide fuel cells." *Advanced Energy Materials* 8,
26

no. 11 (2018) 1702708. https://doi.org/10.1002/aenm.201702708

[17] J. A. Kilner, M. Burriel. "Materials for intermediate-temperature solid-oxide fuel cells." *Annual Review of Materials Research* 44 (2014) 365-393.

https://doi.org/10.1146/annurev-matsci-070813-113426

[18] Z. S. Talaei, H. Salamati, A. Pakzad. "Fabrication and investigation of electrochemical characterization of Ba based cathodes." *International Journal of Hydrogen Energy* 35, no. 17 (2010) 9401-9404. https://doi.org/10.1016/j.ijhydene.2010.05.078

[19] N. Ortiz-Vitoriano, C. Bernuy-Lopez, I. R. de Larramendi, R. Knibbe, K. Thydén, A. Hauch, P. Holtappels, T. Rojo. "Optimizing solid oxide fuel cell cathode processing route for intermediate temperature operation." *Applied Energy* 104 (2013) 984-991.

https://doi.org/10.1016/j.apenergy.2012.12.003

[20] G. Cheng-Zhi, J. Zhou, H. L. Bao, C. Peng, X. Lin, G. P. Xiao, J. Q. Wang, Z. Y. Zhu. "Study of the relationship between the local geometric structure and the stability of $La_{0.6}Sr_{0.4}MnO_{3-\delta}$ and $La_{0.6}Sr_{0.4}FeO_{3-\delta}$ electrodes." *Nuclear Science and Techniques* 30.2 (2019) 21.

https://doi.org/10.1007/s41365-019-0550-1

[21] X. Lu, X. Yang, L. Jia, B. Chi, J. Pu, J. Li. "First principles study on the oxygen reduction reaction of the $La_{1-x}Sr_xMnO_{3-\delta}$ coated $Ba_{1-x}Sr_xCo_{1-y}Fe_yO_{3-\delta}$ cathode for solid oxide fuel cells." *International Journal of Hydrogen Energy* 44, no. 31 (2019) 16359-16367.

https://doi.org/10.1016/j.ijhydene.2019.04.271

[22] K. Huang, J. B. Goodenough. *Solid oxide fuel cell technology: principles, performance and operations*. Elsevier, 2009.

[23] J. Chen, D. Wan, X. Sun, B. Li, M. Lu. "Electrochemical impedance spectroscopic characterization

of impregnated La$_{0.6}$Sr$_{0.4}$Co$_{0.2}$Fe$_{0.8}$O$_{3-\delta}$ cathode for intermediate-temperature SOFCs." *International Journal of Hydrogen Energy* 43, no. 20 (2018) 9770-9776.

https://doi.org/10.1016/j.ijhydene.2018.03.223

[24] A. Mroziński, S. Molin, J. Karczewski, T. Miruszewski, P. Jasiński. "Electrochemical properties of porous Sr$_{0.86}$Ti$_{0.65}$Fe$_{0.35}$O$_3$ oxygen electrodes in solid oxide cells: Impedance study of symmetrical electrodes." *International Journal of Hydrogen Energy* 44, no. 3 (2019) 1827-1838.

https://doi.org/10.1016/j.ijhydene.2018.11.203

[25] X-D. Zhou, J. B. Yang, E-C. Thomsen, Q. Cai, B. J. Scarfino, Z. Nie, G. W. Coffey, W. J. James, W. B. Yelon, H. U. Anderson, L. R. Pederson. "Electrical, Thermoelectric, and Structural Properties of La(M$_x$Fe$_{1-x}$)O$_3$ (M=Mn, Ni, Cu)." *Journal of the Electrochemical Society* 153, no. 12 (2006) J133-J138. https://doi.org/10.1149/1.2358840

[26] L. Kindermann, D. Das, D. Bahadur, R. Wei, H. Nickel, K. Hilpert. "Chemical Interactions between La-Sr-Mn-Fe-O-Based Perovskites and Yttria-Stabilized Zirconia." *Journal of the American Ceramic Society* 80, no. 4 (1997) 909-914. https://doi.org/10.1111/j.1151-2916.1997.tb02921.x

[27] L. Kindermann, D. Das, H. Nickel, K. Hilpert. "Chemical compatibility of the LaFeO3 base perovskites (La$_{0.6}$Sr$_{0.4}$)$_z$Fe$_{0.8}$M$_{0.2}$O$_{3-\delta}$ (z=1, 0.9; M=Cr, Mn, Co, Ni) with yttria stabilized zirconia." *Solid State Ionics* 89, no. 3-4 (1996) 215-220. https://doi.org/10.1016/0167-2738(96)00366-9

[28] L. Mikkelsen, I. G. K. Andersen, E. M. Skou. "Oxygen transport in La$_{1-x}$Sr$_x$Fe$_{1-y}$Mn$_y$O$_{3-\delta}$ perovskites." *Solid State Ionics* 152 (2002) 703-707.

https://doi.org/10.1016/S0167-2738(02)00412-5

[29] J. A. Lane, S. J. Benson, D. Waller, J .A .Kilner. "Oxygen transport in La$_{0.6}$Sr$_{0.4}$Co$_{0.2}$Fe$_{0.8}$O$_{3-\delta}$." *Solid State Ionics* 121.1-4 (1999) 201-208. https://doi.org/10.1016/S0167-2738(99)00014-4

[30] I. Yasuda, K. Ogasawara, M. Hishinuma, T. Kawada, M. Dokiyab. "Oxygen tracer diffusion coefficient of (La, Sr) MnO$_{3\pm\delta}$." *Solid State Ionics* 86 (1996) 1197-1201. https://doi.org/10.1016/0167-2738(96)00287-1

[31] Y. Wang, L. Zhang, C. Xia. "Enhancing oxygen surface exchange coefficients of strontium-doped lanthanum manganates with electrolytes." *International Journal of Hydrogen Energy* 37.3 (2012) 2182-2186. https://doi.org/10.1016/j.ijhydene.2011.11.008

[32] L. Ronghui, D. Qingshan, M. Wenhui, W. Hua, Y. Bin, D. Yongnian, M. Xueju. "Preparation and characterization of component materials for intermediate temperature solid oxide fuel cell by glycine-nitrate process." *Journal of Rare Earths* 24, no. 1 (2006) 98-103.

https://doi.org/10.1016/S1002-0721(07)60333-0

[33] Y. S. Chung, T. Kim, T. H. Shin, H. Yoon, S. Park, N. M. Sammes, W. B. Kim, J. S. Chung. "In situ preparation of a La$_{1.2}$Sr$_{0.8}$Mn$_{0.4}$Fe$_{0.6}$O$_4$ Ruddlesden–Popper phase with exsolved Fe nanoparticles as an anode for SOFCs." *Journal of Materials Chemistry A* 5, no. 14 (2017) 6437-6446. https://doi.org/10.1039/C6TA09692A

[34] F. Barbir, *PEM fuel cells: theory and practice*. Academic Press, 2012.

[35] K. Chen, N. Li, N. Ai, M. Li, Y. Cheng, W. D. A. Rickard, J. Li, S. P. Jiang. "Direct application of cobaltite-based perovskite cathodes on the yttria-stabilized zirconia electrolyte for intermediate temperature solid oxide fuel cells." *Journal of Materials Chemistry A* 4, no. 45 (2016) 17678-17685. https://doi.org/10.1039/C6TA07067A

[36] K. Kakinuma, S. Machida, T. Arisaka, H. Yamamura, T. Atake. "Cathodic characteristics of (La$_{0.6}$Sr$_{0.4}$)(Mn$_{1-x}$Fe$_x$)O$_{3-\delta}$ for a solid oxide fuel cell with a (Ba$_{0.3}$Sr$_{0.2}$La$_{0.5}$) InO$_{2.75}$ electrolyte" *Solid State Ionics* 176, no. 31-34 (2005) 2405-2410. https://doi.org/10.1016/j.ssi.2005.04.048

[37] A. A. Samat, A. A. Jais, M. R. Somalu, N. Osman, A. Muchtar, K. L. Lim. "Electrical and electrochemical characteristics of La$_{0.6}$Sr$_{0.4}$CoO$_{3-\delta}$ cathode materials synthesized by a modified

citrate-EDTA sol-gel method assisted with activated carbon for proton-conducting solid oxide fuel cell application." *Journal of Sol-Gel Science and Technology* 86, no. 3 (2018) 617-630.

https://doi.org/10.1007/s10971-018-4675-1

[38] J. Wang, Z. Yang, X. He, Y. Chen, Z. Lei, S. Peng. "Effect of humidity on $La_{0.4}Sr_{0.6}Co_{0.2}Fe_{0.7}Nb_{0.1}O_{3-\delta}$ cathode of solid oxide fuel cells." *International Journal of Hydrogen Energy* 44, no. 5 (2019) 3055-3062. https://doi.org/10.1016/j.ijhydene.2018.10.193

[39] A. Zomorrodian, H. Salamati, Z. Lu, X. Chen, N. Wu, A. Ignatiev. "Electrical conductivity of epitaxial $La_{0.6}Sr_{0.4}Co_{0.2}Fe_{0.8}O_{3-\delta}$ thin films grown by pulsed laser deposition." *International Journal of Hydrogen Energy* 35, no. 22 (2010) 12443-12448. https://doi.org/10.1016/j.ijhydene.2010.08.100

[40] A. Pakzad, H. Salamati, P. Kameli, Z. Talaei. "Preparation and investigation of electrical and electrochemical properties of lanthanum-based cathode for solid oxide fuel cell." *International Journal of Hydrogen Energy* 35, no. 17 (2010) 9398-9400.

https://doi.org/10.1016/j.ijhydene.2010.04.163

[41] P. Sohrabi, S. Daneshmandi, H. Salamati, M. Ranjbar. "Pulsed laser deposition of $La_{0.6}Ca_{0.4}Fe_{0.8}Ni_{0.2}O_{3-\delta}$ thin films on $SrTiO_3$: Preparation, characterization and electrical properties." *Thin Solid Films* 571 (2014) 180-186. https://doi.org/10.1016/j.tsf.2014.10.074

[42] B. C. H. Steele. "Survey of materials selection for ceramic fuel cells II. Cathodes and anodes." *Solid State Ionics* 86 (1996): 1223-1234. https://doi.org/10.1016/0167-2738(96)00291-3

[43] S. B. Adler, J. A. Lane, B. C. H. Steele. "Electrode kinetics of porous mixed-conducting oxygen electrodes." *Journal of the Electrochemical Society* 143, no. 11 (1996) 3554-3564.

https://doi.org/10.1149/1.1837252

[44] A. Esquirol, N. P. Brandon, J. A. Kilner, M. Mogensen. "Electrochemical characterization of

La$_{0.6}$Sr$_{0.4}$Co$_{0.2}$Fe$_{0.8}$O$_3$ cathodes for intermediate-temperature SOFCs." *Journal of The Electrochemical Society* 151, no. 11 (2004) A1847-A1855. https://doi.org/10.1149/1.1799391

[45] Z. Pan, Q. Liu, R. Lyu, P. Li, S. H. Chan. "Effect of La$_{0.6}$Sr$_{0.4}$Co$_{0.2}$Fe$_{0.8}$O$_{3-\delta}$ air electrode–electrolyte interface on the short-term stability under high-current electrolysis in solid oxide electrolyzer cells." *Journal of Power Sources* 378 (2018) 571-578. https://doi.org/10.1016/j.jpowsour.2018.01.002

[46] Y. Zheng, C. Zhang, R. Ran, R. Cai, Z. Shao, D. Farrusseng. "A new symmetric solid-oxide fuel cell with La$_{0.8}$Sr$_{0.2}$Sc$_{0.2}$Mn$_{0.8}$O$_{3-\delta}$ perovskite oxide as both the anode and cathode." *Acta Materialia* 57, no. 4 (2009) 1165-1175. https://doi.org/10.1016/j.actamat.2008.10.047

[47] F. H. van Heuveln, H. J. M. Bouwmeester. "Electrode properties of Sr-Doped LaMnO$_3$ on yttria-stabilized Zirconia II. electrode kinetics." *Journal of The Electrochemical Society* 144.1 (1997) 134-140. https://doi.org/10.1149/1.1837375

[48] P. Costamagna, C. Sanna, A. Campodonico, E. M. Sala, R. Sazinas, P. Holtappels. "Electrochemical Impedance Spectroscopy of Electrospun La$_{0.6}$Sr$_{0.4}$Co$_{0.2}$Fe$_{0.8}$O$_{3-\delta}$ Nanorod Cathodes for Intermediate Temperature–Solid Oxide Fuel Cells." *Fuel Cells* 19, no. 4 (2019) 472-483. https://doi.org/10.1002/fuce.201800205

[49] L. Sun, H. Li, J. Zhao, G. Wang, L. Huo, H. Zhao "Effects of calcium doping to oxygen reduction activity on Pr$_{2-x}$Ca$_x$NiMnO6 cathode." *Journal of Alloys and Compounds* 777 (2019) 1319-1326. https://doi.org/10.1016/j.jallcom.2018.11.046

[50] Y. Chen, Y. Bu, Y. Zhang, R. Yan, D. Ding, B. Zhao, S. Yoo, D. Dang, R. Hu, C. Yang, M. Liu. "A highly efficient and robust nanofiber cathode for solid oxide fuel cells." *Advanced Energy Materials* 7, no. 6 (2017) 1601890. https://doi.org/10.1002/aenm.201601890

32
[50]  E. Matte, G. Holzlechner, L. Epple, D. Stolten, P. Lupetina "Impact of silicate substrate and cosintering on cathode performance in an inert substrate-supported solid oxide fuel cell." *Journal of Power Sources* 413 (2019) 334-343. https://doi.org/10.1016/j.jpowsour.2018.12.025

[51]  F. Grimm, N.H. Menzler, P. Lupetin, O. Guillon "Screening of Cathode Materials for Inert-Substrate-Supported Solid Oxide Fuel Cells." *ECS Transactions* 91.1 (2019) 1397-1411. https://doi.org/10.1149/09101.1397ecst